\documentclass{PoS}
\usepackage{epsfig,multicol}
\usepackage{amssymb,amsmath,bm, mathrsfs}
 \usepackage{amsfonts}
\usepackage{amstext}
\usepackage{epsf}
\usepackage{graphicx}
\usepackage{longtable}
\usepackage{afterpage}
\usepackage{placeins}
\usepackage{color}
\usepackage{bbm}
\usepackage{afterpage}

\usepackage{slashed}
\usepackage{textcomp}

\pdfoutput=1

\newcommand{\gev}{\, {\rm GeV}}

\allowdisplaybreaks[2]
\newcommand{\be}{\begin{equation}}
\newcommand{\ee}{\end{equation}}
\newcommand{\bea}{\begin{eqnarray}}
\newcommand{\eea}{\end{eqnarray}}
\newcommand{\nn}{\nonumber}
\newcommand{\bi}{\begin{itemize}}
\newcommand{\ei}{\end{itemize}}

\definecolor{orange}{rgb}{1,0.4,0.1}

\DeclareMathOperator{\diag}{diag}

\newcommand{\lsim}{
\mathrel{\hbox{\rlap{\hbox{\lower4pt\hbox{$\sim$}}}\hbox{$<$}}}}
\newcommand{\gsim}{
\mathrel{\hbox{\rlap{\hbox{\lower4pt\hbox{$\sim$}}}\hbox{$>$}}}}

\title{Phenomenological aspects of flavoured dark matter}

\ShortTitle{Phenomenological aspects of flavoured dark matter}

\author{\speaker{Monika Blanke}\vspace{1mm}\\%
     CERN Theory Division, CH-1211 Geneva 23, Switzerland\vspace{1mm}\\
Institut f\"ur Theoretische Teilchenphysik, Karlsruhe Institute of Technology,
Engesserstra{\ss}e 7,\\ D-76128 Karlsruhe, Germany\vspace{1mm}\\
Institut f\"ur Kernphysik, 
Karlsruhe Institute of Technology,
Hermann-von-Helmholtz-Platz 1,\\ 76344 Eggenstein-Leopoldshafen, Germany\vspace{1mm}\\
        E-mail: \email{monika.blanke@kit.edu}}

\abstract{Flavour symmetries in the dark sector are a theoretically motivated and phenomenologically appealing concept. The dark matter particle can be stabilised with the help of flavour symmetries, without the need to introduce an additional discrete symmetry by hand. Apart from the usual searches in direct and indirect detection experiments and high energy colliders, flavoured dark matter generally also gives rise to new flavour violating interactions leading to interesting signatures in rare meson decays. This proceedings article reviews a simplified model of flavoured dark matter in which the dark matter coupling to quarks constitutes a new source of flavour violation, so that the model goes beyond Minimal Flavour Violation. Particular emphasis is put on the discussion of its phenomenological implications in flavour, collider and direct detection experiments.}

\FullConference{The European Physical Society Conference on High Energy Physics\\
                 22-29 July 2015\\
                 Vienna, Austria}

\begin{document}

\section{Introduction}

Astrophysical and cosmological data tell us that the visible baryonic matter constitutes only a small fraction of the total energy density in the universe. About five times more dark matter (DM) exists which manifests itself through its gravitational interactions with ordinary Standard Model (SM) matter. 

Yet we have very little information on the particle nature of DM. Motivated by the concept of Grand Unification, DM, similarly to the visible matter,  could come in multiple generations differing only in their masses. This idea of flavoured DM has become quite popular in recent years (see \cite{Kile:2013ola} for a review). However most studies so far adopted the Minimal Flavour Violation (MFV) assumption \cite{Buras:2000dm,D'Ambrosio:2002ex} for the dark sector, i.\,e.\ that no new sources of flavour violation are present beyond the SM Yukawa couplings, thus saving well-measured flavour changing neutral current (FCNC) observables from dangerously large new contributions. At the same time however it prevents us from observing DM effects in precision flavour experiments.

\begin{figure}[h!]
\centering
\includegraphics[width=0.3\textwidth]{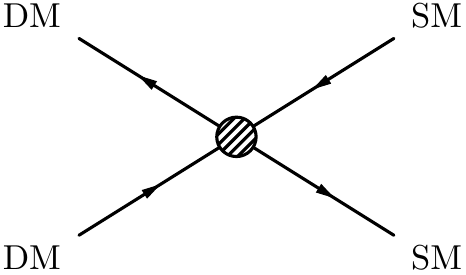}\qquad\qquad\qquad
\includegraphics[width=0.3\textwidth]{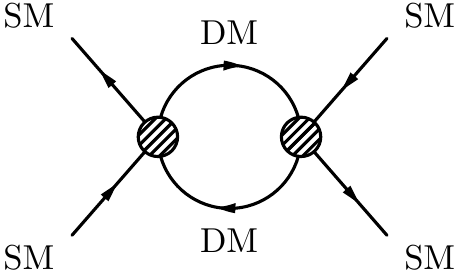}
\caption{Schematic diagrams contributing to experimental constraints on flavoured DM.}
\label{fig:DMloop}
\end{figure}

Our analysis \cite{Agrawal:2014aoa} therefore abandoned the MFV principle and instead presented a simplified model of flavoured DM, in which the coupling of DM to SM quarks constitutes a new source of flavour and CP violation. In addition to the usual DM discovery channels in direct and indirect detection experiments and collider searches, c.\,f.\ left diagram in figure \ref{fig:DMloop}, sizeable contributions to FCNC observables are then also expected, as depicted in the right diagram in figure \ref{fig:DMloop}.

\section{Dark Minimal Flavour Violation}

In the simplified model of \cite{Agrawal:2014aoa} the gauge singlet Dirac-fermionic DM $\chi$ transforms under the triplet representation of a new global flavour symmetry $U(3)_\chi$. Its coupling to the right-handed down type quarks of the SM is mediated by a coloured scalar field $\phi$. The Lagrangian of this simplified model reads
\begin{eqnarray}
  {\cal L}&=& \mathcal{L}_\text{SM}+
  i \bar \chi \slashed{\partial} \chi
  - m_{\chi}  \bar \chi \chi  - (\lambda_{ij} \bar {d_{R}}_i \chi_j \phi + {\rm h.c.}) \nn \\
  && \qquad\,
  +
  (D_{\mu} \phi)^{\dagger} (D^{\mu} \phi) - m_{\phi}^2 \phi^{\dagger} \phi
  +\lambda_{H \phi}\, \phi^{\dagger} \phi\, H^{\dagger} H
  +\lambda_{\phi\phi}\, \phi^{\dagger} \phi\, \phi^{\dagger} \phi \,,\label{eq:Lagrangian}
\end{eqnarray}
whith $i,j=1,2,3$ being flavour indices.

The new Yukawa coupling $\lambda$ is assumed to be the only new source of flavour symmetry breaking, in addition to the SM Yukawa couplings $Y_{u,d}$. Conceptually this ansatz extends the MFV principle to the dark sector flavour group, and is therefore named {\it Dark Minimal Flavour Violation (DMFV)}. Despite its conceptual analogy to the MFV hypothesis, the phenomenology is very different, as the new flavour violating coupling $\lambda$ is unrelated to the SM Yukawas. Therefore DMFV can lead to large new physics contributions to FCNC observables.

The DMFV hypothesis has a number of interesting implications. First, the mass splittings among the dark flavours are generated by non-universalities in the coupling matrix $\lambda$:
\be\label{eq:mchi}
  m_{\chi_i}
  =
  m_{\chi} (\mathbbm{1}  +\eta\, \lambda^\dagger \lambda+\dots)_{ii}\,.
\ee

Second,  the DMFV assumption significantly reduces the number of free parameters. A convenient parametrisation is given by
\be
\lambda = U_\lambda D_\lambda\,,
\ee
where $U_\lambda$ is a unitary matrix containing three mixing angles $\theta_{ij}^\lambda$ and three complex phases $\delta^\lambda_{ij}$ ($ij=12,13,23$) parametrised as in \cite{Blanke:2006xr}. $D_\lambda$ is a real and diagonal matrix, for which we use the parametrisations
\be\label{eq:Dlambda}
D_\lambda \equiv \diag(D_{\lambda,11},D_{\lambda,22},D_{\lambda,33})=
\lambda_0\cdot \mathbbm{1}+\diag(\lambda_1,\lambda_2,-(\lambda_1+\lambda_2))\,.
\ee

Last but not least the DMFV hypothesis ensures DM stability, in complete analogy to the stability of DM in  the MFV case \cite{Batell:2011tc}. The flavour
symmetry $U(3)_Q \times U(3)_u \times U(3)_d \times U(3)_\chi$
broken only by the Yukawa couplings
$Y_u$, $Y_d$ and $\lambda$, together with $SU(3)_\text{QCD}$ implies an
unbroken $\mathbbm{Z}_3$ symmetry, under which only the new particles
$\chi_i$ and $\phi$ are charged. 

%Before moving on we note that the model described by the Lagrangian \eqref{eq:Lagrangian} is the minimal mocel realising the DMFV assumption. We hence refer to it as the {\it minimal DMFV (mDMFV) model}. While for the subsequent phenomenological discussion we restrict our attention to the mDMFV model for simplicity, we note that many conclusions hold more generally within the DMFV framework.

\section{Precision flavour constraints}

\begin{figure}[h!]
\centering
\includegraphics[width=0.33\textwidth]{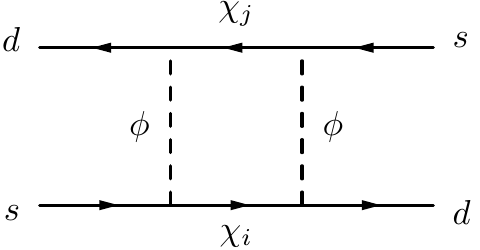}
\caption{New contribution to $K^0-\bar K^0$ mixing in DMFV.}
\label{fig:kkbar}
\end{figure}

New flavour violating interactions at the weak scale are strongly constrained by meson mixing observables. In DMFV new contributions to $K^0-\bar K^0$ mixing are generated by the box diagram in figure \ref{fig:kkbar}. Analogous diagrams contribute to $B_{d,s}-\bar B_{d,s}$ mixing.

The DMFV contributions to the $K-\bar K$, $B_d-\bar B_d$ and $B_s-\bar B_s$ mixing amplitudes are schematically given by
\be
M_{12}^{K,\text{new}} \sim
  \left((\lambda\lambda^\dagger)_{sd}\right)^2 { F(x) }\,,\qquad 
M_{12}^{d,\text{new}} \sim
  \left((\lambda\lambda^\dagger)_{bd}\right)^2 { F(x) }\,,\qquad 
M_{12}^{s,\text{new}} \sim
  \left((\lambda\lambda^\dagger)_{bs}\right)^2 { F(x) }\,,
\ee
where $F(x)$ is the relevant loop function that is flavour independent. We see that the size of new contributions to FCNCs is determined by the off-diagonal entries of $(\lambda\lambda^\dagger)$ which, in order not to spoil the good agreement with the data, have to be small.

\begin{figure}[h!]
\centering
\includegraphics[width=.5\textwidth]{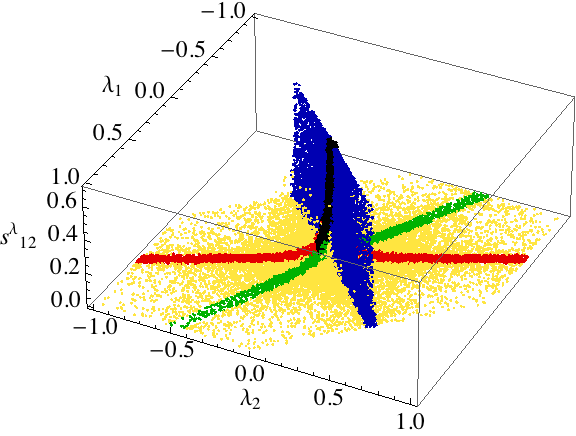}
\caption{\label{fig:scen}Allowed ranges for the mixing angle $s^\lambda_{12}=\sin\theta^\lambda_{12}$ as a function of the non-universality parameters $\lambda_1$ and $\lambda_2$. The five scenarios are: universality (black), small mixing (yellow), 12-degeneracy (blue), 13-degeneracy (red), 23-degeneracy (green).}
\end{figure}

Such a structure can be achieved in various ways, as discussed in detail and checked numerically in \cite{Agrawal:2014aoa}. In short, if a sizeable mixing angle $\theta^\lambda_{ij}$ is present, then the entries $D_{\lambda,ii}$ and $D_{\lambda,jj}$ of the diagonal component of $\lambda$
have to be quasi degenerate. Consequently we end up with the five scenarios for the structure of $\lambda$ pointed out in \cite{Agrawal:2014aoa} and depicted in figure \ref{fig:scen}:
\begin{enumerate}
\item Universality scenario -- The diagonal coupling matrix $D_\lambda$ is nearly universal and large mixing angles in $U_\lambda$ are allowed.
\item Small mixing scenario  -- The mixing matrix $U_\lambda$ is close to the unit matrix, while $D_\lambda$ is arbitrary.
\item $ij$-degeneracy scenarios ($ij=12,13,23$) -- Two elements of $D_\lambda$ are almost equal, $D_{\lambda,ii}\simeq D_{\lambda,jj}$, and only the corresponding mixing angle $s^\lambda_{ij}$ is allowed to be large.
\end{enumerate}

The new contributions to rare $K$ and $B$ decays, to electroweak precision observables, and to electric dipole moments on the other hand are small and do not constrain DMFV further.

\section{Dark matter phenomenology}

Flavour physics gives us valuable information on the structure of the coupling matrix $\lambda$, but it cannot constrain the mass spectrum $m_{\chi_i}$ of the three dark generations. However there are good reasons to assume that the lightest state is $b$-flavoured, i.\,e.\ it couples dominantly to the bottom quark. In this case the agreement with direct detection experiments and LHC searches can be improved. Additionally, $b$ flavoured DM produces interesting $b$-jet signatures at the LHC and can explain the excess $\gamma$-rays observed at the galactic centre \cite{Agrawal:2014una}.

We also require the DM to be a thermal relic and assume that the mass splitting between the DM and the heavier flavours is sizeable, $\gsim 10\%$. Then the heavier states decay before the time of freeze-out so that the observed relic abundance translates into a relation between the coupling $D_{\lambda,33}$, 
 the DM and the mediator mass. The quasi-degenerate case has also been considered in \cite{Agrawal:2014aoa}.

\begin{figure}[h!]
  \includegraphics[width=0.3\textwidth]{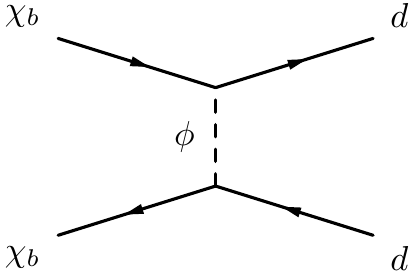}\hfill
  \includegraphics[width=0.3\textwidth]{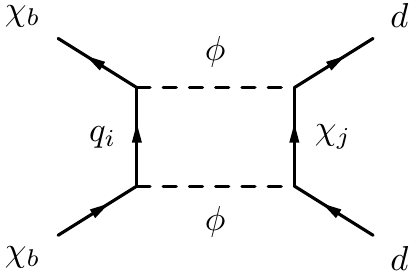}\hfill
  \includegraphics[width=0.3\textwidth]{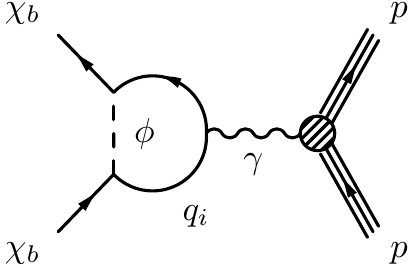}
  \caption{Diagrams contributing to WIMP-nucleon scattering in the mDMFV model.}
  \label{fig:dd}
\end{figure}

Several diagrams yield relevant contributions to direct DM detection, see figure \ref{fig:dd}. 
Note that with the precision achieved by present experiments, mainly LUX \cite{Akerib:2013tjd}, we have become sensitive to DM quark scattering at the one loop level. Thus even if the tree level diagram is suppressed by a small mixing angle $\theta^\lambda_{13}$, large  contributions stemming from box and penguin diagrams are still present and need to be controlled. 

In fact the one loop box and photon penguin diagrams shown in figure \ref{fig:dd} can not be suppressed by small flavour mixing angles. Luckily however these two contributions have opposite signs, so that for a certain range of the coupling $D_{\lambda,11}$, their cancellation becomes effective. DMFV is therefore a natural realisation of xenophobic DM \cite{Feng:2013fyw}.

\begin{figure}[h!]
\centering
\includegraphics[width=.49\textwidth]{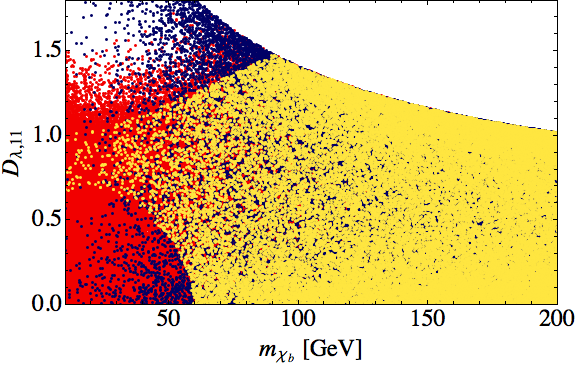}
\caption{\label{fig:mchi-D11}First generation coupling $D_{\lambda,11}$ as functions of the DM mass $m_{\chi_b}$. The red points satisfy the bound from LUX, while the blue points satisfy the flavour constraints. For the yellow points both LUX and flavour constraints are imposed.}
\end{figure}

In figure \ref{fig:mchi-D11} one can see that for large DM masses $m_{\chi_b}\gsim 100\gev$ only the thermal relic constraint restricts the allowed range for $D_{\lambda,11}$. However for smaller DM masses a non-trivial interplay between the LUX and flavour constraints can be observed which restricts
 $D_{\lambda,11}$ to lie in a certain range. 

\section{LHC signatures}

The new particles introduced in DMFV are expected to be accessible to the LHC. Due to the conserved $\mathbb{Z}_3$ symmetry they have to be pair produced, and the lightest state (assumed to be $\chi_b$) escapes detection and leads to missing energy signatures. The heavier $\chi_i$ states, due to the small mass splitting, give rise to soft jets or photons when decaying into the lightest state.

\begin{figure}[!h]
  \begin{center}
    \includegraphics[width=0.45\textwidth]{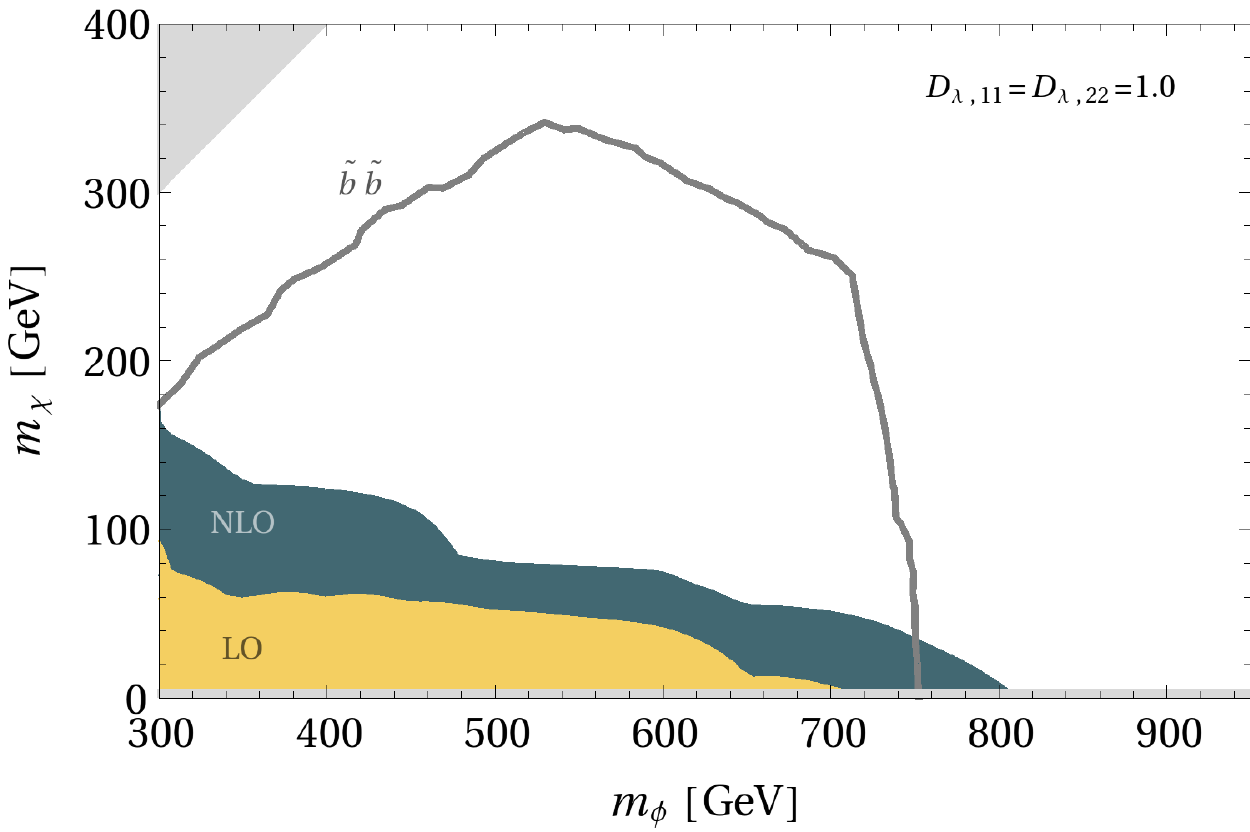}
    \quad
    \includegraphics[width=0.45\textwidth]{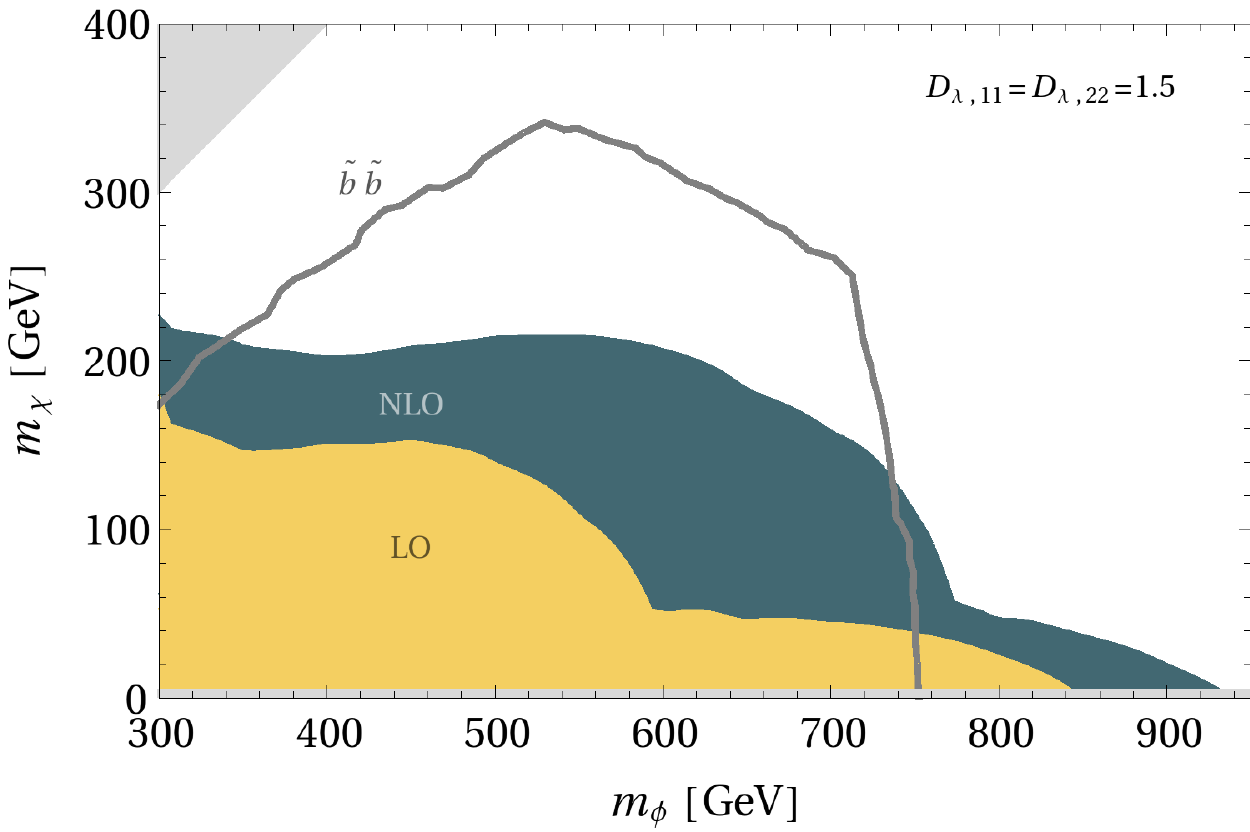}
  \end{center}
  \caption{Limits on DMFV from the CMS 19.5
  fb$^{-1}$ sbottom~\cite{CMS:2014nia} and
  squark~\cite{Chatrchyan:2014lfa} search.  The DM coupling to $b$-quark, $D_{\lambda,33}$
  is fixed everywhere by the corresponding relic abundance
  constraint, and all mixing angles are set to zero for simplicity.}
  \label{fig:sbottoms1}
\end{figure}

The signatures of DMFV are very similar to the ones from popular SUSY models with $R$-parity conservation. The production of $\phi$ pairs is constrained by sbottom and light squark searches, leading to the same final state signatures. Therefore the DMFV exclusion contours in the $m_\chi$-$m_\phi$ plane ,shown in figure \ref{fig:sbottoms1}, can be obtained by adapting the  CMS 19.5
  fb$^{-1}$ sbottom~\cite{CMS:2014nia} and   squark~\cite{Chatrchyan:2014lfa} searches, by taking into account both the changes in production cross section and branching ratios. For light DM masses these bounds reach up to mediator masses of $m_\phi \simeq 850\gev$, however they become significantly weaker for increasing $m_\chi$ mass.

\begin{figure}[!h]
  \centering
    \includegraphics[width=0.45\textwidth]{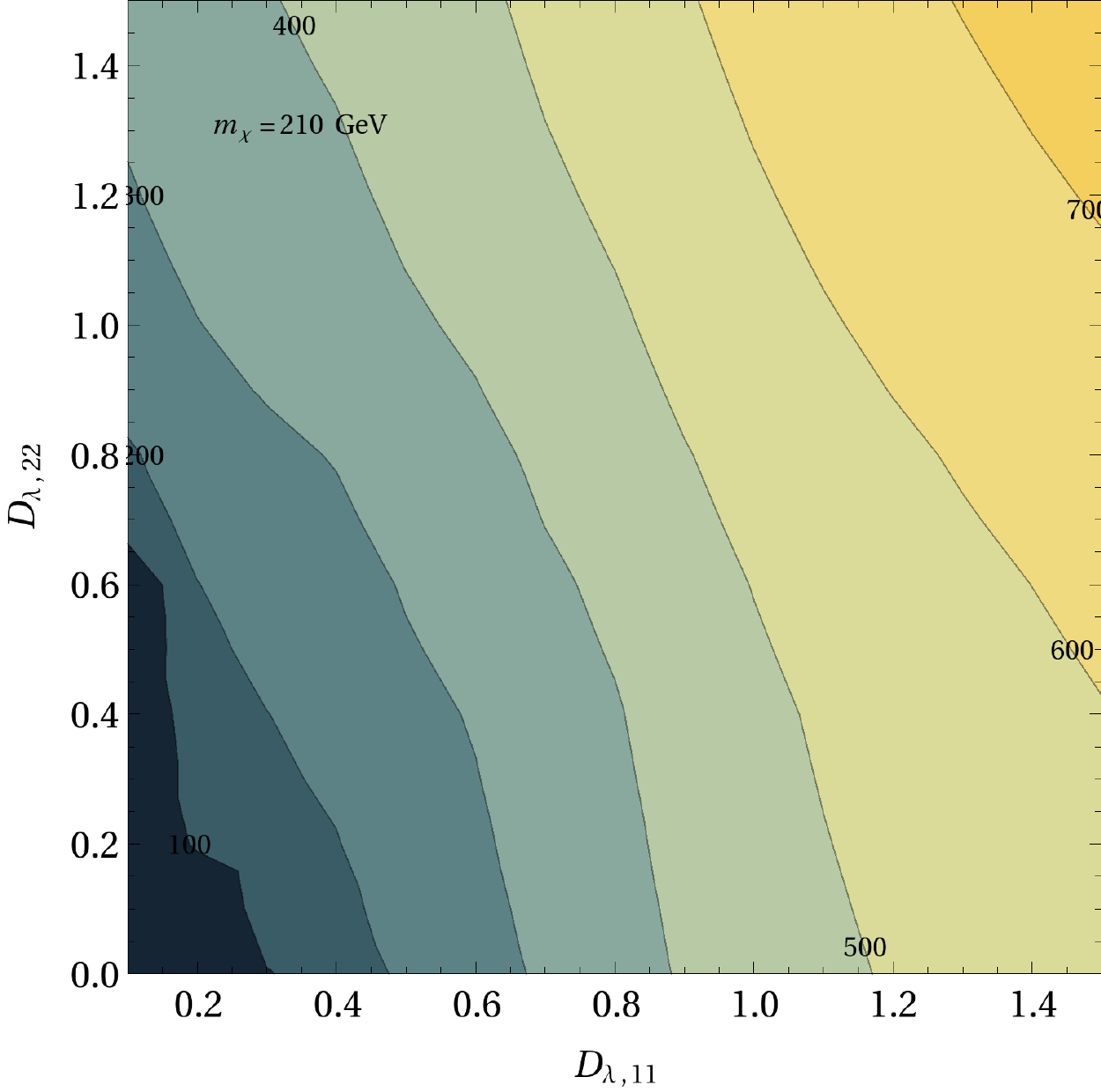}
    \caption{Limits on DMFV in terms of mass of $\phi$ (in
  GeV) from the CMS monojet
  search~\cite{CMS:rwa}. }
  \label{fig:monojetscoup}
\end{figure}

Monojet searches instead are sensitive to direct DM pair production, as well as to pair production of the heavier $\chi$ flavours. They provide constraints on the mediator mass $m_\phi$ as function of the couplings $D_{\lambda,ii}$, shown in figure \ref{fig:monojetscoup}. Contrary to the dijet constraints, in this case the obtained result is rather insensitive to the value of the DM mass $m_\chi$.

DMFV also gives rise to some interesting signatures accessible to future searches. Similarly to the case of flavour violating squark decays \cite{Blanke:2013uia}, $\phi$ pair production and decay can lead to one $b$-jet and one light jet in the final state, accompanied by missing energy. Due to the stringent flavour physics constraints, such a signature is unlikely to emerge in SUSY models.

\section{Summary}

Motivated by  grand unification scenarios it is theoretically well-motivated to consider that DM, in analogy to the SM fermions, carries flavour quantum number and comes in multiple copies.
Such setups lead to interesting phenomenological implications, in particular when new flavour violating interactions with the SM quarks are present. Flavour precision data provide useful constraints on the structure of the DM coupling matrix, while the mass spectrum can be constrained mostly from direct DM and LHC searches. A non-trivial interplay of the various sectors is identified, proving the importance of studying the various sectors in a correlated manner. 

{\it I would like to thank Prateek Agrawal and Katrin Gemmler for a very enjoyable and fruitful collaboration on the topic presented here.}

\end{document}